\documentclass[twocolumn,showpacs,preprintnumbers,amsmath,amssymb,subfig,aps,prb]{revtex4}

\usepackage{graphicx}
\usepackage{hhline}
\usepackage{xcolor}

\begin{document}

\title{Classical magnetization of a four-dimensional Platonic solid}

\author{N. P. Konstantinidis}
\affiliation{Department of Mathematics and Natural Sciences, The American University of Iraq, Sulaimani, Kirkuk Main Road, Sulaymaniyah, Kurdistan Region, Iraq}

\date{\today}

\begin{abstract}
The 600-cell is a regular 4-polytope that is a four-dimensional analog of a Platonic solid. Three-dimensional Platonic solids with icosahedral $I_h$-symmetry have been shown to have a discontinuous ground-state magnetization response in an external field at the classical and quantum level, when spins mounted on their vertices interact according to the antiferromagnetic Heisenberg model. The discontinuities are not due to anisotropy in spin space, but rather to the special connectivity of the molecules. Here the nearest-neighbor antiferromagnetic XX and Heisenberg models in a magnetic field are considered for classical spins mounted on the 120 vertices of the 600-cell. The ground-state magnetization response is rich, characterized by six magnetization discontinuities in the XX case, and six magnetization and three susceptibility discontinuities in the Heisenberg case. This demonstrates that going from three to four spatial dimensions enriches the ground-state magnetization response.
\end{abstract}


\maketitle


The antiferromagnetic XX and Heisenberg models (AXXM and AHM respectively) describe interactions between localized spins in two and three spin-space dimensions respectively
\cite{Auerbach98,Fazekas99}. Of particular interest are lattices and molecules where antiferromagnetic interactions do not support antiparallel nearest-neighbor spins in the classical lowest-energy configuration. This is due to competing interactions and is known as frustration \cite{Lhuillier01,Misguich03,Ramirez05,Schnack10,Florek19,Schmidt20}. Frustration has important consequences both at the classical and quantum level.
For example, the classical ground-state magnetization curve in an external field can exhibit features such as discontinuities in the magnetization and susceptibility, which are absent for bipartite structures.

The icosahedron and the dodecahedron belong to the family of Platonic solids \cite{Plato}, and have icosahedral-$I_h$ point-group symmetry \cite{Altmann94}. In Platonic solids all the vertices and edges are geometrically equivalent. The icosahedron is made of triangles, perhaps the simplest frustrated unit, and the AHM on it has a ground-state magnetization discontinuity in an external field at the classical level \cite{Schroeder05,NPK15}. The dodecahedron consists of pentagons, which are also frustrated, and the AHM on it has ground-state magnetization discontinuities at both the classical and quantum limit \cite{Coffey92,NPK05,NPK07}. It is the smallest fullerene molecule \cite{Fowler95}, and its discontinuous magnetic response is shared by larger fullerenes with $I_h$ symmetry. The two molecules are dual of one another, and they have been shown to possess a similar low-energy spectrum. Fullerenes with different symmetries have also been found to exhibit rich magnetic behavior \cite{NPK09,NPK17,NPK18}, while the next-bigger icosahedral fullerene dual, the pentakis dodecahedron, shows a very rich magnetization response as a function of the relative strength of its two symmetrically unique exchange constants \cite{NPK21}.

In this paper, motivated by the rich magnetic properties of highly symmetrical molecules, the nearest-neighbor AXXM and AHM in a magnetic field on the 600-cell, the four-dimensional analog of the icosahedron, are considered. The 600-cell consists of 120 vertices, 720 edges, and 1200 triangles \cite{Coxeter73}. Each vertex is 12-fold coordinated, and all the vertices and edges are geometrically equivalent. The ground-state magnetization response is very rich, with six magnetization discontinuities for the AXXM, and six magnetization and three susceptibility discontinuities for the AHM. The increased spatial dimensionality and coordination number of the 600-cell leads to a richer magnetization response in comparison with the three-dimensional Platonic solids.


The 600-cell has $N=120$ vertices. It is the dual of the 120-cell, the four-dimensional analog of the dodecahedron. The Hamiltonian for spins $\vec{s}_i$ and $\vec{s}_j$ mounted on the vertices $i,j=1,\dots,N$ of the molecule is

\begin{equation}
H_n = \sum_{<ij>} \sum_{\sigma=1}^n s_i^\sigma s_j^\sigma - h \sum_{i=1}^{N} s_i^z
\label{eqn:model}
\end{equation}

The first term describes exchange interactions, which are two-dimensional in spin space for the AXXM ($n=2$) and three-dimensional for the AHM ($n=3$). The brackets in $<ij>$ indicate that interactions are limited to nearest neighbors. The first term defines the unit of energy. The second term is the energy due to an external magnetic field of strength $h$, taken along one of the interaction axes. The spins $\vec{s}_i$ are unit vectors whose direction is determined by a polar $\theta_i$ and an azimuthal $\phi_i$ angle in three-dimensional spin space, with only the polar angle needed in two dimensions.

The lowest-energy configuration of Hamiltonian (\ref{eqn:model}) is a result of the competition for minimization between the exchange and the magnetic energy, with the frustrated connectivity of the 600-cell playing a central role. Minimization of the Hamiltonian gives the lowest-energy spin configuration as a function of $h$ \cite{Coffey92,NPK07,NPK13,NPK15,NPK15-1,NPK16,NPK16-1,NPK17,NPK17-1,NPK18,NPK21,Machens13,NPK22}.



\begin{figure}
\includegraphics[width=3.5in,height=2.5in]{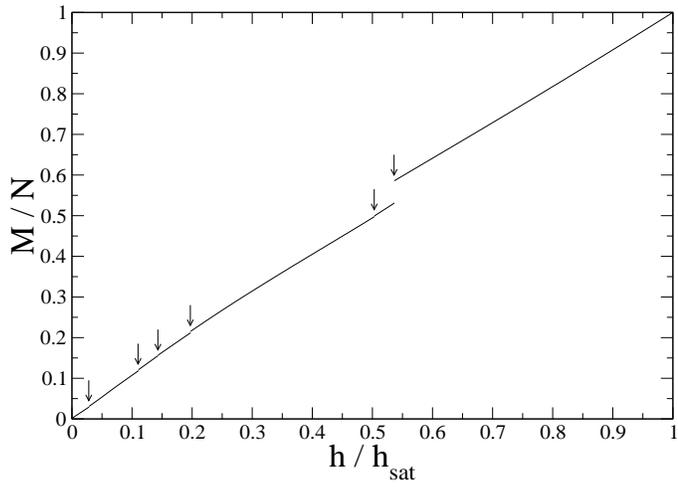}
\vspace{0pt}
\caption{Ground-state magnetization per spin $\frac{M}{N}$ in the ground state of the AXXM model (\ref{eqn:model}) as a function of the magnetic field $h$ over its saturation value $h_{sat}$. The solid arrows point to the locations of the magnetization discontinuities.
}
\label{fig:600-cell-AXXM}
\end{figure}

In the absence of magnetic field the ground-state energy
per bond of the AXXM equals
-0.2710231 and the magnetization
per spin $1.436124 \times 10^{-3}$. The magnetization as a function of $h$ in the lowest-energy configuration of Hamiltonian (\ref{eqn:model}) has six discontinuities. Table \ref{table:classicalmagndiscAXXM} lists the fields for which they occur, the magnetization right below and above the jump and the corresponding change. Except for the highest-field jump the rest are associated with smaller changes in the magnetization. The magnetization as a function of the external field is shown in Fig. \ref{fig:600-cell-AXXM}.

\begin{figure}
\includegraphics[width=3.5in,height=2.5in]{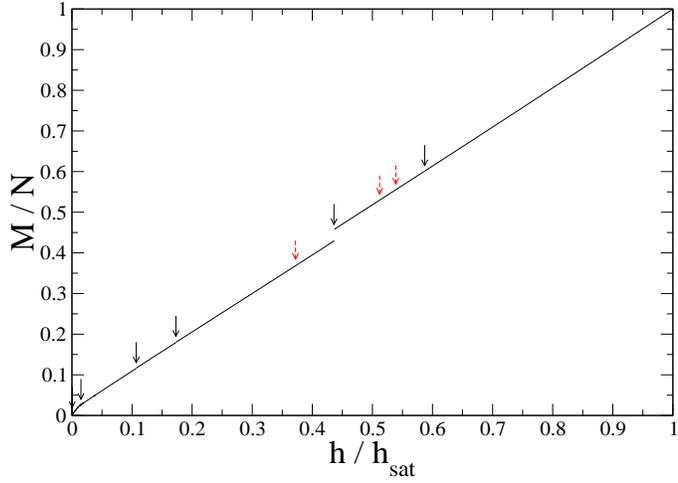}
\vspace{0pt}
\caption{Ground-state magnetization per spin $\frac{M}{N}$ in the ground state of the AHM model (\ref{eqn:model}) as a function of the magnetic field $h$ over its saturation value $h_{sat}$. The solid arrows point to the locations of the magnetization discontinuities and the dashed arrows to the locations of the susceptibility discontinuities.
}
\label{fig:600-cell-AHM}
\end{figure}

In the case of the AHM, the zero-field ground-state energy
per bond is lowered to
-0.2909842 \cite{NPK22} and the magnetization per spin is zero. The magnetization as a function of $h$ in the lowest-energy configuration of Hamiltonian (\ref{eqn:model}) has a total of six magnetization and three susceptibility discontinuities, listed in Tables \ref{table:classicalmagndiscAHM} and \ref{table:classicalsuscdiscAHM}. Again one of the high-field jumps is associated with a significantly higher magnetization change than the rest. The magnetization as a function of the external field is shown in Fig. \ref{fig:600-cell-AHM}. The response is much richer than the one of the three-dimensional icosahedron, which has a single magnetization discontinuity in an external field.


Platonic solids are unique in that they have equivalent vertices and are made up of a single type of polygon. Here the lowest-energy configuration magnetic response of a four-dimensional Platonic solid, the 600-cell, has been calculated for the AXXM and the AHM. It has a significant number of magnetization and susceptibility discontinuities, showing that going from three to four-dimensional Platonic solids enriches the magnetic properties.



\bibliography{sixhundredcell}

\newpage


\begin{table}
\begin{center}
\caption{Magnetization discontinuities of the AXXM (\ref{eqn:model}). The columns give the value of the magnetic field $h$ for which the discontinuity appears over its saturation value $h_{sat}$, the reduced magnetization $\frac{M}{N}$ below and above the discontinuity, and the reduced magnetization change. The saturation magnetic field $h_{sat}=15.70820$.}
\begin{tabular}{c|c|c|c}
$\frac{h}{h_{sat}}$ & $({\frac{M}{N}})_{below}$ & $({\frac{M}{N}})_{above}$ & $\frac{\Delta M}{N} (\times 10^{-3})$ \\
\hline
 0.0284380 & 0.0292860 & 0.0307879 & 1.50188 \\
\hline
 0.1108123 & 0.1187076 & 0.1211165 & 2.408849 \\
\hline
 0.1433832 & 0.1557494 & 0.1569121 & 1.16263488 \\
 \hline
 0.1974717 & 0.2119367 & 0.2153885 & 3.45184 \\
\hline
 0.5034437 & 0.4977158 & 0.4995004 & 1.78462 \\
\hline
 0.5364572 & 0.5311187 & 0.5858516 & 54.732929
\end{tabular}
\label{table:classicalmagndiscAXXM}
\end{center}
\end{table}


\begin{table}
\begin{center}
\caption{Magnetization discontinuities of the AHM (\ref{eqn:model}). The columns give the value of the magnetic field $h$ for which the discontinuity appears over its saturation value $h_{sat}$, the reduced magnetization $\frac{M}{N}$ below and above the discontinuity, and the reduced magnetization change. The saturation magnetic field $h_{sat}=15.70820$.}
\begin{tabular}{c|c|c|c}
$\frac{h}{h_{sat}}$ & $({\frac{M}{N}})_{below}$ & $({\frac{M}{N}})_{above}$ & $\frac{\Delta M}{N} (\times 10^{-4})$ \\
\hline
 0.0005047 & 0.0015516 & 0.001682 & 1.30 \\
\hline
 0.0150123 & 0.0266205 & 0.0266332 & 0.127 \\
\hline
 0.1072130 & 0.1160844 & 0.1167255 & 6.4108 \\
\hline
 0.1729125 & 0.1793218 & 0.1798761 & 5.5432 \\
\hline
 0.4365934 & 0.4298667 & 0.4579936 & 281.26869 \\
\hline
 0.5870964 & 0.6012727 & 0.6012773 & 0.0466
\end{tabular}
\label{table:classicalmagndiscAHM}
\end{center}
\end{table}

\begin{table}
\begin{center}
\caption{Susceptibility discontinuities of the AHM (\ref{eqn:model}). The columns give the value of the magnetic field $h$ for which the discontinuity appears over its saturation value $h_{sat}$, and the reduced magnetization $\frac{M}{N}$ at this field value. The saturation magnetic field $h_{sat}=15.70820$.}
\begin{tabular}{c|c}
 $\frac{h}{h_{sat}}$ & $\frac{M}{N}$ \\
\hline
 0.37245 & 0.36872 \\
\hline
 0.51263 &  0.53026 \\
\hline
 0.53929 &  0.55568
\end{tabular}
\label{table:classicalsuscdiscAHM}
\end{center}
\end{table}

\end{document}